# Cu NQR study of host and guest spin dynamics in impurity-doped YBa$_2$Cu$_4$O$_8$

Yutaka Itoh, Takato Machi, Nobuaki Watanabe, and Naoki Koshizuka

Superconductivity Research Laboratory, International Superconductivity Technology Center, 10-13 Shinonome 1-chome, Koto-ku, Tokyo 135-0062, Japan

We present the planar $^{63}$Cu nuclear spin-lattice relaxation study of impurity-doped YBa$_2$(Cu$_{1-x}$M$_x$)$_4$O$_8$ (M=Ni, Zn) at $T$=4.2-300 K with Cu nuclear quadrupole resonance (NQR) spin-echo technique. In the light of an impurity-induced NQR relaxation theory, we estimated the host Cu nuclear spin-lattice relaxation time $(T_1)_{HOST}$, the impurity-induced Cu nuclear spin-lattice relaxation time ($\tau_1$), and the wipeout region around the impurity in the superconducting state. The pseudo spin-gap behavior of the host Cu antiferromagnetic spin fluctuation above $T_c$ is not readily removed by a small amount of the impurity. The Zn-induced non-Korringa behavior of $(T_1)_{HOST}$ below $T_c$ is in sharp contrast to the doping effect of the magnetic impurity Ni.

## 1. Introduction

NMR/NQR technique has supplied microscopic information on the impurity-doped high-$T_c$ superconductors. The $^{89}$Y NMR study for Zn-doped YBa$_2$Cu$_3$O$_{7-}$ is a widely known example, which demonstrates the appearance of Zn-induced satellite signal with a Curie law (~$1/T$) [1]. The planar Cu NMR/NQR technique can measure the strongly enhanced antiferromagnetic spin susceptibility $\chi''(q{\sim}Q, \omega)$ (the wave vector $q$, $Q$=[$\pi$, $\pi$] and the frequency $\omega$), characteristic of the CuO$_2$ plane in the high-$T_c$ superconductor [2]. However, the recovery curve $p(t) \equiv 1-M(t)/M(\infty)$ of the planar Cu(2) nuclear magnetization $M(t)$ for the impurity-doped material is not a simple exponential function with a single $T_1$ [3]. How to estimate $T_1$ has been a long term problem. It has been highly desired to estimate separately host Cu NQR spin-lattice relaxation time $(T_1)_{HOST}$ and impurity-induced Cu NQR relaxation time $\tau_1$.

In this paper, we present the planar $^{63}$Cu nuclear spin-lattice relaxation study of the impurity-doped YBa$_2$(Cu$_{1-x}$M$_x$)$_4$O$_8$ (M=Ni, Zn) in the light of an impurity-induced NQR relaxation theory with a wipeout effect [4]. Here, Ni is a magnetic impurity (3$d^8$, $S$=1), and Zn is a nonmagnetic impurity (3$d^{10}$, $S$=0; spinless defect). We have estimated separately the host $(T_1)_{HOST}$ and the impurity-induced $\tau_1$ from the analysis of the nonexponential recovery curves.

## 2. Experimental

Powder samples YBa$_2$(Cu$_{1-x}$M$_x$)$_4$O$_8$ (M=Ni, $x$=0, 0.009, 0.02, 0.03; M=Zn, $x$=0, 0.005, and 0.010) were prepared by a solid-state reaction method and a high-oxygen-pressure technique with hot isostatic pressing apparatus [5, 6]. The values of the impurity content $x_{ICP}$ were estimated by inductively coupled plasma atomic emission spectroscopy (ICP), which are somewhat different from the nominal $x$ reported in refs. [5-7]. The sharp superconducting transitions were observed at $T_c$=66, 44, 15 K for Ni doping ($x_{ICP}$=0.009, 0.02, 0.03) and at $T_c$=82, 68, 56 K for Zn doping ($x_{ICP}$=0, 0.005, 0.010), from *dc* magnetization measurements, indicating the homogeneous distribution of Ni and Zn. Figure 1 shows $T_c$ as a function of $x_{ICP}$ for Ni or Zn doping. From measurements of polycrystalline resistivity [5, 6], the superconductor-to-semiconductor transition at low temperature was observed around the critical impurity concentration of $x_{ICP}$=0.03~0.04. Zero-field Cu NQR measurements were carried out with a coherent-type pulsed spectrometer at $T$=4.2-300 K. Nuclear spin-lattice relaxation was measured by an inversion recovery spin-echo technique, where the $^{63}$Cu(2) nuclear spin-echo intensity $M(t)$ was recorded as a function of the time $t$ after an inversion pulse.

## 3. Wipeout number in the superconducting state: Ni versus Zn

Figure 2 shows the experimental recovery curves $p(t)$ for (a)Ni doping ($x_{ICP}$=0.009) and for (b)Zn doping ($x_{ICP}$=0.01) at 4.2 K. The solid curves are the least-squares fits of nonexponential function of,

$$p(t) = p(0)\exp[-\frac{3t}{T_{1\,HOST}} - N_c \left( e^{-3t/t_c} - 1 + \sqrt{\frac{3\pi t}{t_c}} erf\sqrt{\frac{3t}{t_c}} \right)], \quad (1)$$

where $p(0)$ is a fraction of the initially inverted magnetization, $(T_1)_{HOST}$ is the Cu NQR relaxation time due to the host Cu electron spin fluctuation, $N_c$ (0 ≤ $N_c$ ≤ 1) is the wipeout number per unit volume, $t_c(=r_c^{2d}/C_{im})$ is the impurity-induced Cu nuclear spin-lattice relaxation time $T_1(r)(=r^{2d}/C_{im})$ at the exclusion radius $r_c$ ($r$ is a distance from an impurity to the nuclear site, $d$ the space dimension of nuclear sites, and $C_{im}$ an impurity magnetic auto-correlation function), and erf is the error function [4, 8]. The fit parameters are $p(0)$, $(T_1)_{HOST}$, $N_c$ and $t_c$.

The nuclei within the radius $r_c$ are out of NQR and then removed from the consideration. $N_c$ is given by $x_{plane}S_d r_c^d/V$, where $x_{plane}$ is an in-plane impurity concentration, $V$ is a unit volume (=$a^2$ with the in-plane lattice constant $a$, if $d$=2), and $S_d = 2\pi^{d/2}/d\Gamma(d/2)$ with the gamma function Γ is the surface area of a unit sphere in $d$ dimension. The wipeout number $N_c$ below $T_c$ could not be estimated from the integrated intensity of Cu NQR spectra, because of the Meissner-Ochsenfeld shielding effect, even on the powder. But, our relaxation analysis with eq. (1) can work well even below $T_c$ if $N_c$ is a fitting parameter.

The stretched exponential function with the curly bracket of eq. (1) is derived from the random superposition of $T_1(r)(=r^{2d}/C_{im})$ process locally enhanced around the impurity. The enhanced $T_1(r)$ around Ni results from the classical dipole-dipole coupling ($r^{-3}S_zI_\pm$ with Ni impurity spin $S$ and Cu nuclear spin $I$) and/or the RKKY coupling ($\sim r^{-d}$), whereas the origin of $T_1(r)$ around Zn is not clear. The Zn-induced local moments [1] and staggered moments [9, 10] are the promising candidates. In either cases, eq. (1) was derived.

The estimated values of $(T_1)_{HOST}$, $N_C$ and $t_c$ at 4.2 K are given in Figs. 2(a) and 2(b). The number of wipeout cells is estimated to be about 10~20 per Zn ($x_{plane}$=0.01~0.02) but ~1 per Ni ($x_{plane}$~0.01). If $N_C = (r_c/a)^2$ (circle), $(r_c/a)^2$ (square) and $4r_c/a$ (diagonal), then $r_c/a$=2~3, 3~4, and 3~5 are obtained for Zn doping. This is nearly the same as the antiferromagnetic correlation length $\xi_{AF}/a$=3~3.5 [11]. Thus, we found larger wiepout region around Zn than that around Ni in the superconducting states. On the left-hand side in Fig. 2, the schematic illustration shows the electronic state of the $CuO_2$ plane with Ni or Zn, deduced from the analysis with eq. (1). The shaded area in the circle indicates the wipeout region around Zn. For simplicity, the electronic density oscillations around the impurities are omitted. The wipeout region around Zn can remind us of a magnetic vortex or a staggered Skyrmion with a radius $r_c$ [12] rather than the charge-spin stripe formation. The magnetic vortex or Skyrmion can also act as a strong pair-breaking effect.

## 4. Host and impurity-induced Cu spin dynamics: Ni versus Zn

Although eq. (1) with the finite $N_c$ reproduces the Zn-induced nonexponential recovery curve better than that without $N_c$, there is a shortcoming of the overestimation of $N_c$ above $T_c$, partly because the mean impurity spacing $r_{imp} \sim (V/x_{plane}S_d)^{1/d}$ is assumed to be $r_{imp}$ [8]. Hence, we adopt the next best policy of the fixed $N_c$=0, and fit the following equation (2) to the experimental recovery curves, in order to extract the temperature dependences of the host Cu NQR relaxation rate $(1/T_1)_{HOST}$ and of the impurity-induced Cu relaxation rate $1/\tau_1 (= N_c^2/t_c)$,

$$p(t) = p(0) \exp\left[-\frac{3t}{T_1}\bigg|_{HOST} - \sqrt{\frac{3t}{\tau_1}}\right], \quad (2)$$

which is derived from eq. (1) in the limit of $r_c \to 0$ ($1/t_c$ and $N_c \to 0$) [4]. The fit parameters are $p(0)$, $(T_1)_{HOST}$ and $\tau_1$. The fitted curves based on eq. (2) can be seen in refs. [7, 13].

Figure 3 shows the impurity effect on the temperature dependence of the estimated relaxation rates $1/\tau_1$ and $(1/T_1T)_{HOST}$, for Ni doping [(a)$1/\tau_1$ and (b)$(1/T_1T)_{HOST}$] and for Zn doping [(c)$1/\tau_1$ and (d)$(1/T_1T)_{HOST}$]. In Fig. 3(d), we obtain a significant result that the rapid increase of $(1/T_1T)_{HOST}$ (non-Korringa behavior) is induced by a small amount of Zn in the superconducting state. Far below $T_c$, the impurity effect on the host $(1/T_1T)_{HOST}$ is quite different between Ni and Zn, whereas the impurity-induced $1/\tau_1$ levels off, being of the same order of magnitude between Ni and Zn.

In the unitarity limit, the impurity in a $d_{x^2-y^2}$-wave superconductor induces a virtual bound state (resonance state) around the impurity site [14, 15]. In Fig. 3(b), the nearly $T$-independent $(1/T_1T)_{HOST}$ due to Ni doping below 15 K, Korringa-like behavior, can be explained either by the weak enhancement of the Ni-induced antiferromagnetic spin fluctuation in the unitarity limit or by the intermediate strength scattering [16]. In Fig. 3(d), the rapid increase of $(1/T_1T)_{HOST}$ due to Zn doping below $T_C$ indicates the strong enhancement of the Zn-induced antiferromagnetic spin fluctuation in the resonance states in the unitarity limit [16].

Figure 4 shows the impurity doping effect on the pseudo spin-gap behavior of the host Cu NQR relaxation rate $(1/T_1T)_{HOST}$ for Ni and for Zn. In pure $YBa_2Cu_4O_8$, $(1/T_1T)_{HOST}$ decreases below about $T_S$=160 K, which is called the pseudo spin-gap behavior. Clearly, the pseudo spin-gap behavior of $(1/T_1T)_{HOST}$ is robust for a small amount of the impurity doping; either Ni ($x_{ICP}$=0.009, 0.02, 0.03) or Zn ($x_{ICP}$=0.005, 0.01). This is sharply contrast to the result in ref. [17] but consistent with the conclusion of ref. [10]. The robust pseudo spin-gap behavior of $(1/T_1T)_{HOST}$ is also consistent with the robust pseudogap in the $^{89}$Y NMR [1], the $^{63}$Cu NMR [10], inelastic neutron scattering [18], and in-plane electrical resistivity [19]. If the antiferromagnetic spin fluctuation is the dynamical stripe fluctuation and if the impurity acts as a pinning site, then the spin fluctuation energy $\Gamma(Q)$ is reduced by softening effect, which leads to the divergence of $(1/T_1T)_{HOST}$( $1/\Gamma(Q)$) at low temperature. The observed robust behavior of $(1/T_1T)_{HOST}$ indicates that a small amount of the impurity does not act as a pinning site.

## 5. Summary

The wipeout number $N_C$ in the superconducting state is estimated from the planar Cu nuclear spin-lattice relaxation curve. The observed nonexponential recovery curves indicate the inhomogeneous spin dynamics in real space, where $N_C$ around Zn is much larger than that around Ni. We found that the pseudo spin-gap temperature $T_S$ of the host Cu $(1/T_1T)_{HOST}$ is not readily decreased by a small amount of Zn nor of Ni, and that the non-Korringa behavior induced by Zn below $T_C$ suggests the presence of the Zn-induced enhanced antiferromagnetic correlation in a $d_{x^2-y^2}$-wave superconductivity, in contrast to Ni doping.

## Acknowledgments

This work was supported by New Energy and Industrial Technology Development Organization (NEDO) as Collaborative Research and Development of Fundamental Technologies for Superconductivity Applications.

# Figure Captions

Fig. 1 $T_c$ versus the impurity content $x_{ICP}$. The dashed line is a guide for the eye, indicating the low temperature localization region.

Fig. 2 The nonexponential recovery curves $p(t)$ 1-$M(t)/M(\ )$ of the planar Cu(2) nuclear magnetization $M(t)$ for (a)Ni doping ($x_{ICP}$=0.009) and for (b)Zn doping ($x_{ICP}$=0.01) at 4.2 K. The solid curves are fitted results based on eq. (1). On the left-hand side, the schematic illustration of the electronic state of the $CuO_2$ plane with Ni or with Zn is shown. The arrow denotes the local moment of Ni. The "arrow" represents the enhanced spin correlation in the wipeout region around Zn (shaded circles).

Fig. 3 The relaxation rates $1/\tau_1$ and $(1/T_1T)_{HOST}$ estimated by eq. (2) for Ni doping [(a)$1/\tau_1$ and (b)$(1/T_1T)_{HOST}$] and for Zn doping [(c)$1/\tau_1$ and (d)$(1/T_1T)_{HOST}$]. The dotted lines ( $T^2$ for the $d_{x^2-y^2}$-wave superconductivity) are shown as a guide for the eye.

Fig. 4 The impurity doping effect on the pseudo spin-gap behavior of $(1/T_1T)_{HOST}$ for (a)Ni [7] and for (b)Zn [13].

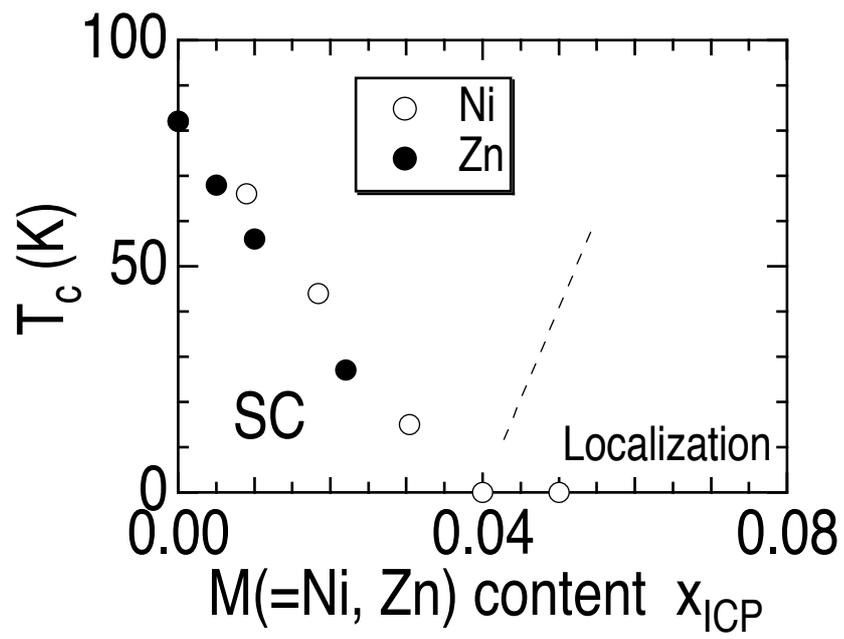

Fig. 1  Y. Itoh et al.

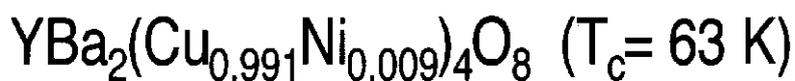

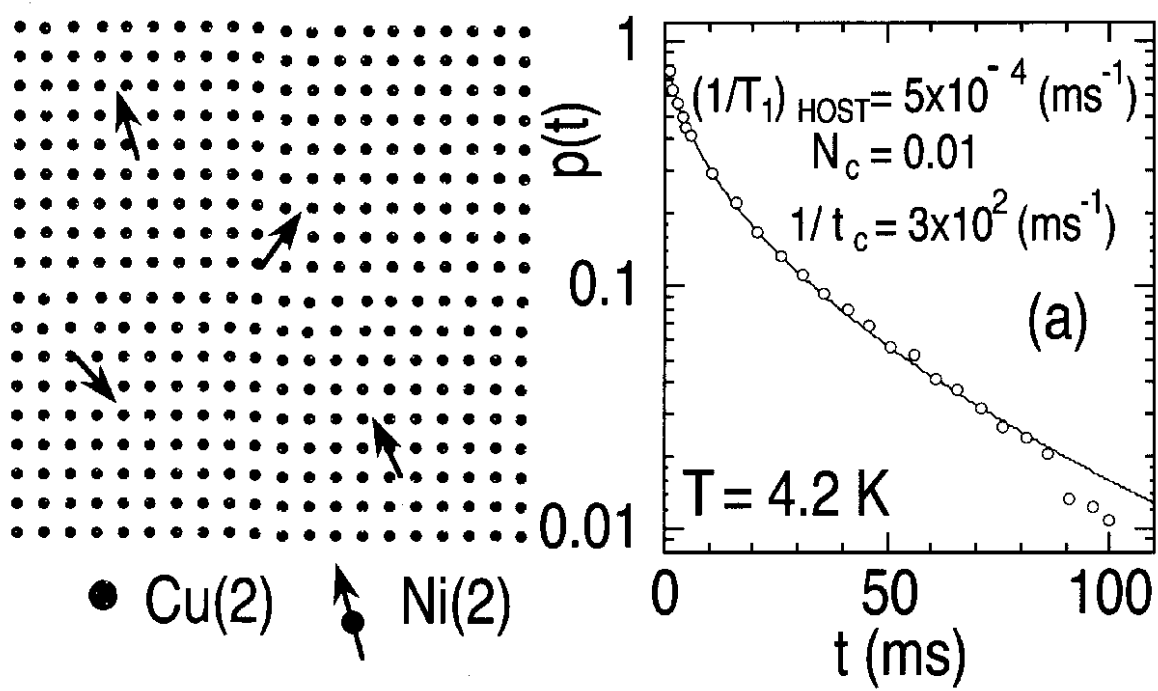

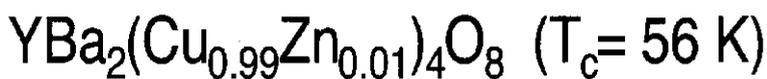

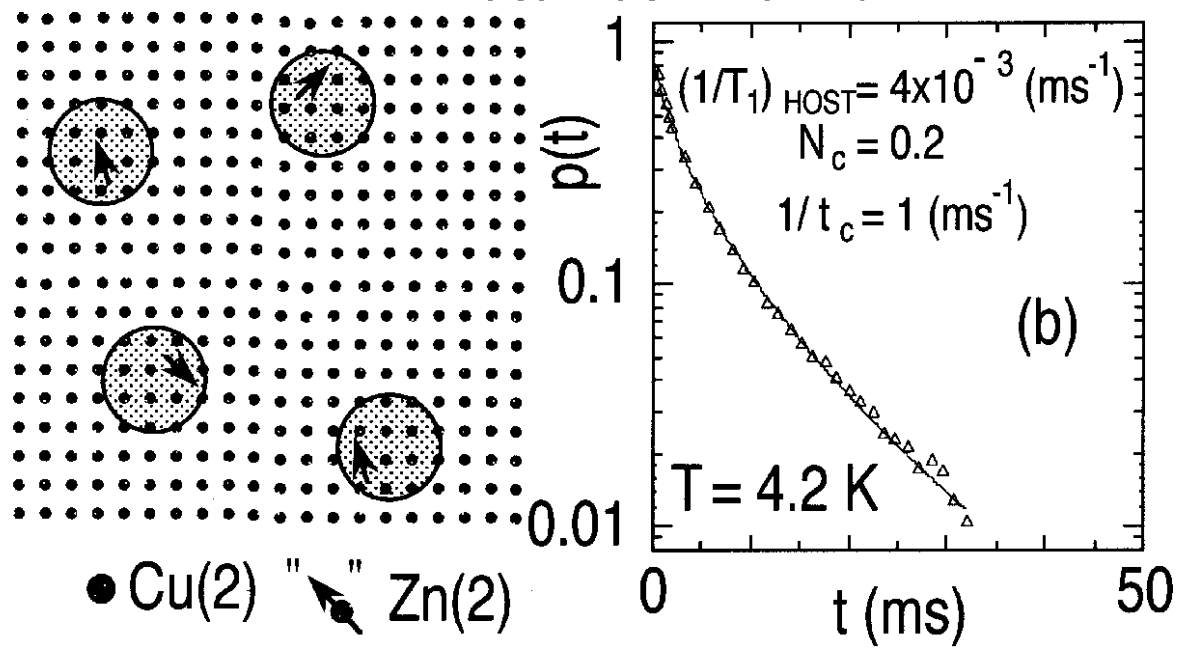

Fig.2    Y. Itoh et al.

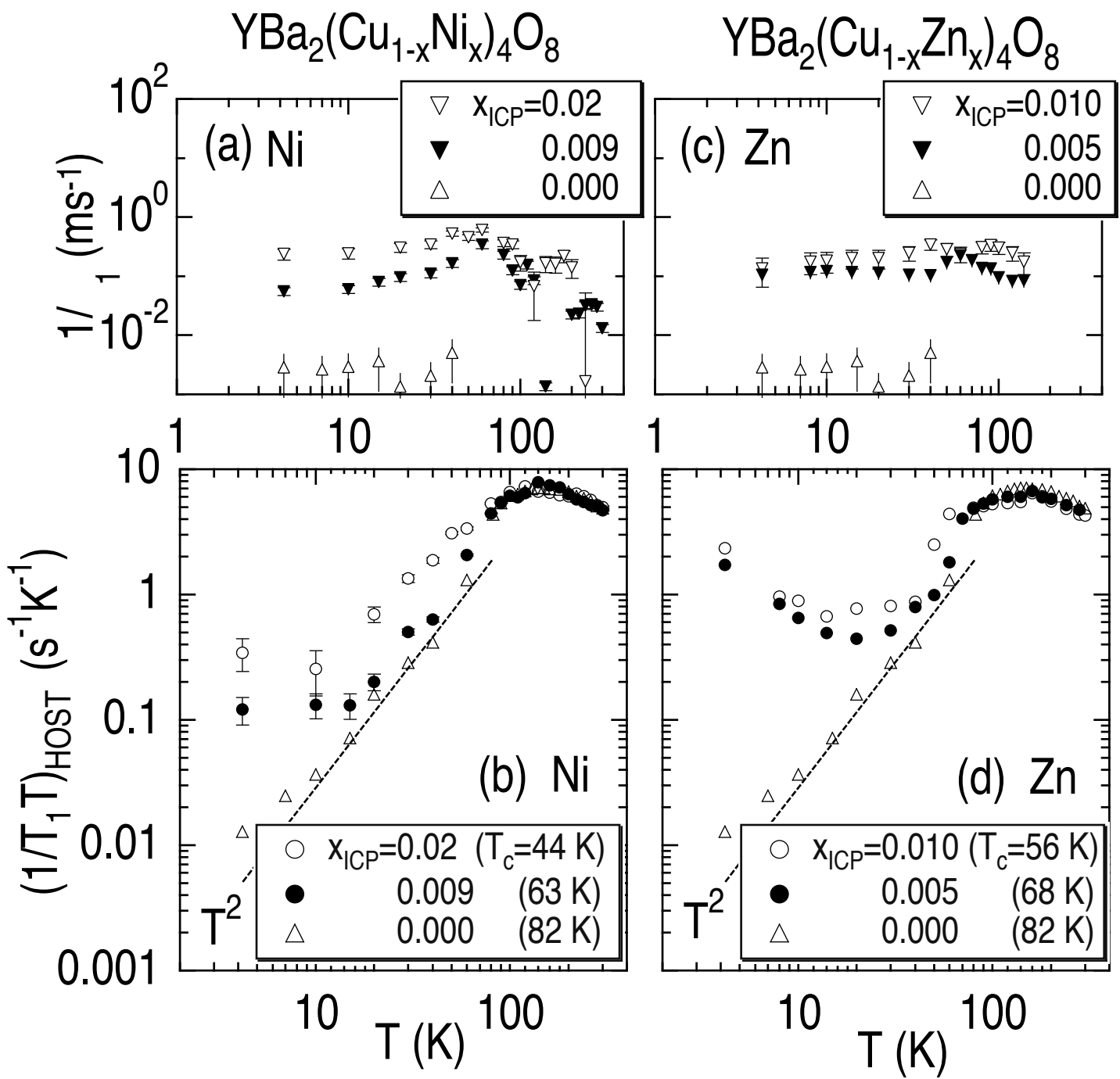

Fig.3 Y. Itoh et al.

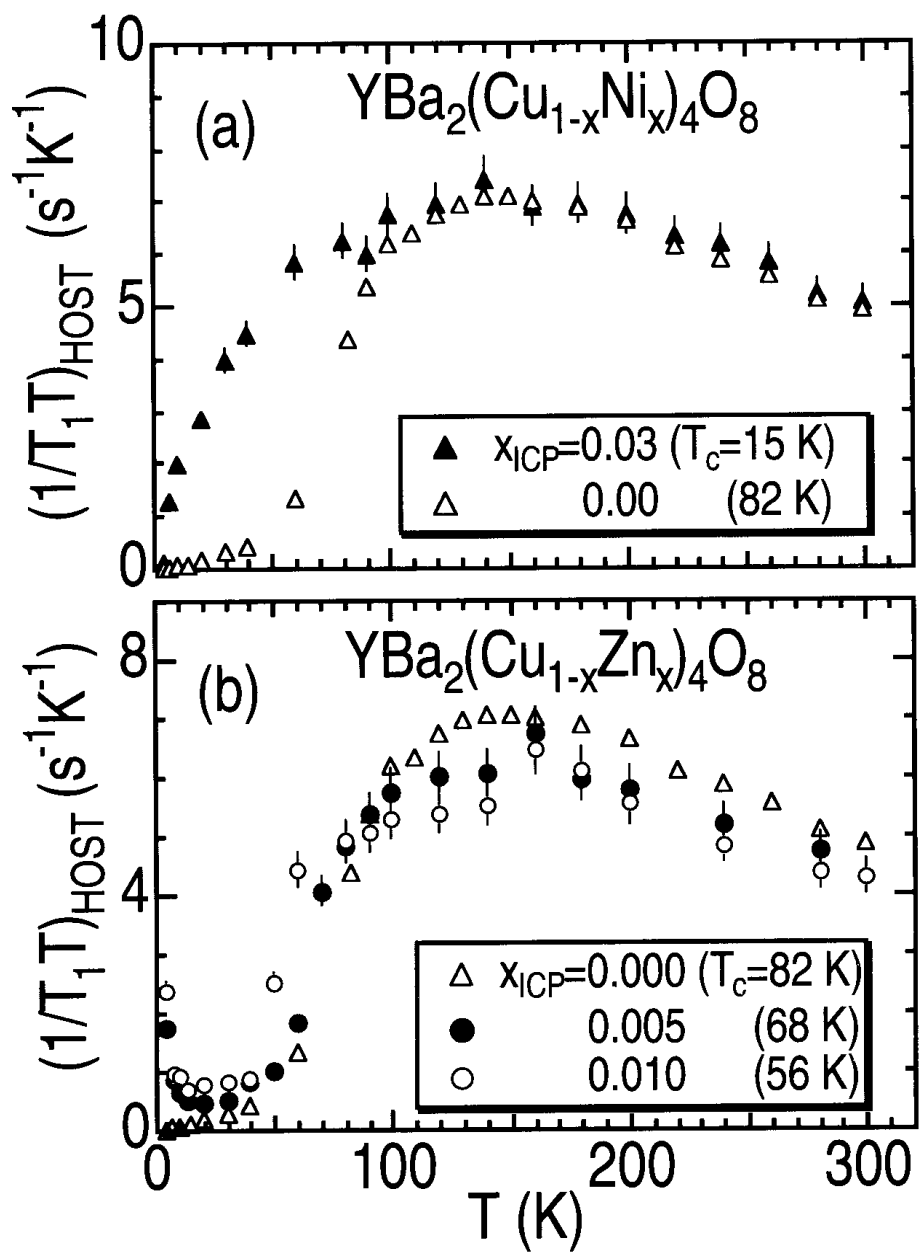

Fig. 4    Y. Itoh et al.